\documentclass[prl,twocolumn,preprintnumbers,amsmath,amssymb]{revtex4}
\usepackage{epsfig,psfrag}

\catcode`\@=11
\def \be  {\begin{equation}}
\def \ee  {\end{equation}}
\def \ba  {\begin{eqnarray}}
\def \ea  {\end{eqnarray}}
\def \baa {\begin{eqnarray*}}
\def \eaa {\end{eqnarray*}}
\def \bb  {\begin {thebibliography} }
\def \eb  {\end{thebibliography}}
\def \lab #1 {\label{#1}}
\newcommand{\ft}[2]{{\textstyle\frac{#1}{#2}}}

\newcommand\re[1]{(\ref{#1})}

\def \matrix #1 {\left(\begin{array}{cc} #1 \end{array}\right)}

\def \Re {\mathop{\rm Re}\nolimits}

\def \e  {\mathop{\rm e}\nolimits}
\def \sech{\mathop{\rm sech}\nolimits}
\newcommand\lr[1]{{\left({#1}\right)}}

\newcommand{\as}{\ifmmode\alpha_{\rm s}\else{$\alpha_{\rm s}$}\fi}
\newcommand{\asbar}{\ifmmode\bar{\alpha}_{\rm s}\else{$\bar{\alpha}_{\rm s}$}\fi}

\font\cmss=cmss12 
\def\inbar{\,\vrule height1.5ex width.4pt depth0pt}
\def\IC{\relax\hbox{$\inbar\kern-.3em{\rm C}$}}
\def\IZ{\relax{\hbox{\cmss Z\kern-.4em Z}}}
\def\IR{{\hbox{{\rm I}\kern-.2em\hbox{\rm R}}}}

\def\IP{{\hbox{{\rm I}\kern-.2em\hbox{\rm P}}}}
\def\II{\hbox{{1}\kern-.25em\hbox{l}}}

\begin{document}
\preprint{LPT-Orsay-07-74}
\title{Cusp anomalous dimension in maximally supersymmetric \\ Yang-Mills theory at strong coupling}
\author{B. Basso$^1$, G.P. Korchemsky$^1$, J. Kota\'{n}ski$^{1,2}$}
\affiliation{$^1$Laboratoire de Physique Th\'eorique, Universit\'e de Paris XI and CNRS (UMR 8627),
91405 Orsay C\'edex, France}
\affiliation{$^2$Institute of Physics, Jagellonian University, Reymonta 4,
PL-30-059 Cracow, Poland}
\begin{abstract}
We construct an analytical solution to the integral equation which is
believed to describe logarithmic growth of the anomalous dimensions of high spin
operators in planar $\mathcal{N}=4$ super~Yang-Mills theory and use it to determine
the strong coupling expansion of the cusp anomalous dimension.

\end{abstract}

\maketitle

{\it 1.~Introduction:} The cusp anomalous dimension is an important observable in
four dimensional gauge theories ranging from QCD to maximally supersymmetric
$\mathcal{N}=4$ Yang-Mills theory (SYM) since it governs the scaling behavior of
various gauge invariant quantities like logarithmic growth of the anomalous
dimensions of high-spin Wilson operators, Sudakov asymptotics of elastic form
factors, the gluon Regge trajectory, infrared singularities of on-shell
scattering amplitudes {\it etc}. By definition~\cite{P80,KR86}, $\Gamma_{\rm
cusp}(g)$ measures the anomalous dimension of a Wilson loop evaluated over a
closed contour with a light-like cusp in Minkowski space-time. It is a function
of the gauge coupling only and its expansion at weak coupling is known in QCD to
three loops~\cite{Vogt04} and in $\mathcal{N}=4$ SYM to four loops~\cite{4-loop}.
Recently, a significant progress has been achieved in determining $\Gamma_{\rm
cusp}(g)$ at strong coupling in planar $\mathcal{N}=4$ SYM. Within the AdS/CFT
correspondence~\cite{Mal97}, $\Gamma_{\rm cusp}(g)$ at strong coupling is related
to the semiclassical expansion of the energy of folded string rotating in
AdS${}_3$ part of the target space~\cite{GPK,FroTse03} (see also~\cite{K02})
\be\label{str}
\Gamma_{\rm cusp}(g) = 2 g - \frac{3\ln 2}{2\pi} + O(1/g)\,,\qquad g =
\frac{\sqrt{\lambda}}{4\pi},
\ee
with $\lambda = g^2_{\rm YM} N_c$ being t' Hooft coupling. On the gauge theory
side, the Bethe ansatz approach to calculating $\Gamma_{\rm cusp}(g)$ in the weak
coupling limit was developed in \cite{Kor95} based on integrability symmetry of
planar Yang-Mills theory to one loop~\cite{BBGK04}. This approach was recently
extended to all loops in planar $\mathcal{N}=4$ SYM. Comparing integrable
structures present on both sides of the AdS/CFT correspondence, an all-loop
asymptotic Bethe ansatz was proposed in \cite{AFS04}. It involves a nontrivial
scattering phase satisfying the crossing symmetry \cite{J06} whose explicit form
was found in \cite{BT05}. This led to an integral equation for the all-loop cusp
anomalous dimension~\cite{BES06,B06}, the Beisert-Eden-Staudacher (BES) equation,
\be\label{BES}
\widehat\sigma (t) = \frac{t}{e^{t}-1}\left[K(2gt, 0) - 4g^{2}\int_{0}^{\infty}
dt' K(2gt, 2gt') \widehat\sigma(t')\right]\! ,
\ee
with $\Gamma_{\rm cusp}(g)=8 g^2 \widehat\sigma (0)$. Here, the kernel $K(t,t')$
is expressed in terms of Bessel functions, $K(t,t')=\sum_{n,m=1}^\infty z_{nm}(g)
J_n(t) J_m(t')/(tt')$, and its explicit form can be found in \cite{BES06}. At
weak coupling, the Neumann series solution to \re{BES} yields perturbative
expansion of $\Gamma_{\rm cusp}(g)$ in powers of $g^2$ which agrees with the
known four-loop result~\cite{4-loop}. At intermediate coupling, the integral
equation \re{BES} was solved numerically in \cite{Benna06}. The obtained solution
for $\Gamma_{\rm cusp}(g)$ was found to be a smooth function of $g$ that matches
for $g>1$ the string theory prediction \re{str} with high accuracy. Moreover, an
exact analytical solution to equation \re{BES} in the limit $g\to \infty$ was
recently constructed in \cite{Alday07} leading to $\Gamma_{\rm cusp}(g) = 2 g +
O(g^0)$, in agreement with \re{str} and with the numerical analysis of
\cite{Benna06}. Equation \re{BES} was analyzed further in \cite{Kostov07} but it
resisted an analytical solution so far. In parallel development, the computation
of the two-loop $O(1/g)$ corrections to the string theory prediction \re{str} was
initiated in \cite{RTT07}. Also, the result \re{str} was reproduced \cite{CK07}
from the quantum string Bethe ansatz for a folded string rotating in
AdS${}_3\times $S${}^1$.

In this Letter, we describe an approach to finding a strong coupling expansion of
the solution to equation \re{BES}. It allows us to determine exact analytical
expressions for the coefficients in the $1/g-$expansion of the cusp anomalous
dimension \re{str} to any desired order.

Let us introduce two even/odd functions $\gamma_\pm(-t)=\pm \gamma_\pm(t)$
\be\label{gamma-def}
\frac{\e^t-1}{t}\widehat\sigma (t) =
\frac{\gamma_+(2gt)}{2gt}+\frac{\gamma_-(2gt)}{2gt}\,.
\ee
Following \cite{Alday07}, we expand $\gamma_\pm(t)$ into the Bessel function
Neumann series
\ba\label{gamma-N}
\gamma_+(t) &=& \mbox{$\sum_{k\ge 1}$} (-1)^{k+1} (2k) J_{2k}(t)\gamma_{2k}\,,
\\[2mm]\nonumber
\gamma_-(t) &=& \mbox{$\sum_{k\ge 1}$} (-1)^{k+1} (2k-1)
J_{2k-1}(t)\gamma_{2k-1}\,,
\ea
with the expansion coefficients $\gamma_k\sim \int_0^\infty dt'/t' J_k(t')
\gamma_\sigma(t')$ ($\sigma=+/-$ for $k=$ even/odd) and sign factors introduced
for the later convenience. Substituting \re{gamma-def} into equation \re{BES} and
separating even/odd in $t$ parts, we find after some algebra that
\re{BES} is equivalent to the (infinite) system of equations %(with $n\ge 1$)
\ba\label{gamma-system}
&& \int_{0}^{\infty}\frac{dt}{t}\bigg[\frac{\gamma_{+}(t)}{1-\e^{-t/2g}}
-\frac{\gamma_{-}(t)}{\e^{t/2g}-1}\bigg]J_{2n}(t) = 0\,,
\\
&&\nonumber \int_{0}^{\infty}\frac{dt}{t}\bigg[\frac{\gamma_{-}(t)}{1-\e^{-t/2g}}
+\frac{\gamma_{+}(t)}{\e^{t/2g}-1}\bigg]J_{2n-1}(t) = \frac{1}{2}\delta_{n, 1}\,,
\ea
with $n\ge 1$. The cusp anomalous dimension can be read from small$-t$ expansion
$\gamma_-(t) = t\, {\Gamma_{\rm cusp}(g)}/{(8 g^2)} + O(t^2)$. At weak coupling,
one finds from \re{gamma-system} that $\gamma_{-}(t)= J_1(t)+O(g^2)$ leading to
${\Gamma_{\rm cusp}(g)}= 4g^2+O(g^4)$ in agreement with the known one-loop
result~\cite{P80,KR86}.

{\it 2.~Exact solution:} The system \re{gamma-system} has the following
remarkable property. Introducing two even/odd functions $\Gamma_{\pm}(t)=
\gamma_{\pm}(t) \mp \gamma_{\mp}(t)\coth{\frac{t}{4g}}$, or equivalently
\be\label{change-gamma}
 2\gamma_{\pm}(t) =
\left[1- \sech{\frac{t}{2g}}\right] \Gamma_{\pm}(t)\pm\tanh{\frac{t}{2g}} \
\Gamma_{\mp}(t)\,,
\ee
we find from \re{gamma-system}
\ba\label{Gamma-system}
&& \int_{0}^{\infty}\frac{dt}{t}\bigg[\Gamma_{+}(t)+\Gamma_{-}(t)\bigg]J_{2n}(t)
= 0\,,
\\\nonumber
&&
\int_{0}^{\infty}\frac{dt}{t}\bigg[\Gamma_{-}(t)-\Gamma_{+}(t)\bigg]J_{2n-1}(t)
=\delta_{n, 1}\,,
\ea
(with $n \ge 1$) and the cusp anomalous dimension is now given by $\Gamma_{\rm
cusp}(g)=-2g \Gamma_+(0)$. Here, in comparison with \re{gamma-system}, the
dependence on $g$ only resides in $\Gamma_\pm(t)$.

At large $g$, we expect from \re{change-gamma} that the functions $\Gamma_\pm(t)$
admit expansion in the Bessel function Neumann series
%Similar to \re{gamma-N}, we look for $\Gamma_\pm(t)$ in the form of the Bessel
%function Neumann series
\ba\label{Gamma-N}
\Gamma_+(t) &=& \mbox{$\sum_{k\ge 0}$} (-1)^{k+1}   J_{2k}(t)\Gamma_{2k}\,,
\\[2mm]\nonumber
\Gamma_-(t) &=& \mbox{$\sum_{k\ge 0}$} (-1)^{k+1} J_{2k-1}(t)\Gamma_{2k-1}\,.
\ea
In distinction to \re{gamma-N}, the first series involves $J_0(t)$ term which
ensures that $\Gamma_+(0)\neq 0$. Indeed, for $t\to 0$ one gets from \re{gamma-N}
and \re{change-gamma} that $\gamma_+(t) \sim t^2$ and $\gamma_-(t)\sim t$ and,
therefore, $\Gamma_+(t)\sim t^0$ and $\Gamma_-(t) \sim t$. Also, in virtue of
$J_{-1}(t)=-J_1(t)$, the coefficient in front of $J_1(t)$ is given by
$(\Gamma_{1}+\Gamma_{-1})$ so that it is only the sum of two coefficients that is
uniquely defined. We make use of this ambiguity to choose $\Gamma_{-1}=1$.

Substitution of \re{Gamma-N} into \re{Gamma-system} yields an infinite system of
finite-difference equations for the coefficients $\Gamma_{k}$. Applying standard
methods, we were able to construct its solution for $\Gamma_{k}$ (with $k\ge -1$)
in the following form (detailed analysis will be published elsewhere)
\be\label{Gamma-exact}
\Gamma_{k} =  -\frac12 \Gamma_{k}^{(0)}+\sum_{p=1}^\infty \frac1{g^p} \left[c_p^-
\Gamma_{k}^{(2p-1)} + c_p^+  \Gamma_{k}^{(2p)} \right]\,,
\ee
where $\Gamma_{k}^{(p)}$ are basis functions independent on $g$
\be\label{Gamma-basis}
\Gamma_{2m}^{(p)} = \frac{\Gamma(m+p-\ft12)}{\Gamma(m+1)\Gamma(\ft12)}, \quad
\Gamma_{2m-1}^{(p)} =
\frac{(-1)^{p}\Gamma(m-\ft12)}{\Gamma(m+1-p)\Gamma(\ft12)}\,,
%= (-1)^p \,\Gamma_{2(k-p)}^{(p)}
\ee
and the expansion coefficients $c_p^\pm$ given by series in inverse powers of the
coupling, $ c_p^\pm = \sum_{r\ge 0} g^{-r} c_{p,r}^\pm$.  The sum over $p$ in the
r.h.s.\ of \re{Gamma-exact} describes the contribution of zero modes of
\re{Gamma-system}. Their dependence on $g$ is fixed by the additional condition
of scaling behavior of $\gamma_\pm(t)$ (see Eqs.\re{scaling} and \re{z-functions}
below). Knowing the $c_p^\pm-$coefficients we can determine the cusp anomalous
dimension $\Gamma_{\rm cusp}(g) = -2g \Gamma_+(0)$ $ = 2g \Gamma_0$ as
\be\label{cusp-c}
\Gamma_{\rm cusp}(g) = {2g}  + \sum_{p=1}^\infty \frac1{g^{p-1}}
\left[\frac{2c_p^-}{\sqrt{\pi}} {\Gamma(2p-\ft32)}\!+\! \frac{2c_p^+}{\sqrt{\pi}}
{\Gamma(2p-\ft12)} \right]\!.
\ee

Let us now establish the relation between the coefficients $\Gamma_n$ and
$\gamma_n$. To this end, we return to the relation \re{change-gamma} and apply
the identities
\ba\label{a}
&& \sech{t}-1 = \mbox{$\sum_{n\ge 1}$}(-1)^n a_{2n}t^{2n}\,,
\\[2mm] \nonumber
&& \tanh{t} = \mbox{$\sum_{n\ge 1}$}(-1)^n a_{2n-1}t^{2n-1}\,,
\ea
where $a-$coefficients with even/odd indices are related to the Euler/Bernoulli
numbers, respectively. This leads to
\ba\label{N}
2\gamma_\pm (t) &=& \sum_{n\ge 1}(-1)^{n+1}
\\ \nonumber
&\times& \bigg[\frac{a_{2n}}{g^{2n}} \lr{{t}/2}^{2n}\Gamma_\pm(t)\mp
\frac{a_{2n-1}}{g^{2n-1}} \lr{{t}/2}^{2n-1}\Gamma_\mp(t)\bigg].
\ea
Replacing $\gamma_\pm(t)$ and $\Gamma_\pm(t)$ by the series \re{gamma-N} and
\re{Gamma-N}, respectively, we make use of the Bessel function Neumann series for
$(t/2)^p J_m(t)$ in the r.h.s.\ of \re{N} to obtain
\ba \label{gamma=Gamma}
&& \!\!\!\!\!\gamma_{2m} =  \sum_{n=1}^m\sum_{j=0}^{m-n+1}\left[ \Gamma_{2j-1}\,
{K_{2m,2j-1}^{2n-1}}
 +\Gamma_{2j}\,  K_{2m,2j}^{2n} \right],
\\ \nonumber
&& \!\!\!\!\!\gamma_{2m-1} = \sum_{n=1}^m\sum_{j=0}^{m-n}\left[ \Gamma_{2j-1}\,
K_{2m-1,2j-1}^{2n} +\Gamma_{2j}\, K_{2m-1,2j}^{2n-1} \right].
\ea
Here the notation was introduced for the coefficients
\ba
&& \hspace*{-5mm} K_{m,j}^n = -\frac{a_n/g^n} {2\Gamma(n)} %\theta\lr{m-n-j}
\\ \nonumber
&& \times\frac{\Gamma(\frac12
(m+j+n))\Gamma(\frac12(m-j+n))}{\Gamma(\frac12(m+j-n)+1)\Gamma(\frac12(m-j-n)+1)}\,,
\ea
which are given by a product of two identical even/odd polynomials of degree $(n
-1)$ in variables $(m\pm j)$
\be\label{K-asym}
K_{m,j}^n = -\frac{2a_n/g^{n}}{4^n\Gamma(n)} \left[\lr{{m+j}}^{\! n-1}\!\!+\ldots
\right]\left[\lr{{m-j}}^{\! n-1}\!\!+\ldots \right],
\ee
with ellipses denoting terms with smaller nonnegative power of $(m\pm j)$.
Replacing $\Gamma_j$ in \re{gamma=Gamma} by their explicit expressions,
Eqs.\re{Gamma-exact} and \re{Gamma-basis}, we express $\gamma_{2m}$ and
$\gamma_{2m-1}$ in terms of yet unknown coefficients $c_p^\pm$. In particular,
the first two terms of their large-$g$ expansion look as
\be\label{example}
\gamma_k = \frac1{g} \gamma_k^{(0)} +\frac1{g^2} \big[ c_-\gamma_k^{(1),-} +
c_+\gamma_k^{(1),+} \big] + O(1/g^{3}),
\ee
with $c_-=c_1^-+\ft12 c_1^+$, $c_+=\ft{16}3 c_1^-+2c_1^+$ and $\gamma_k^{(0,1)}$
given in terms of $\Gamma_{2m}^{(p)}$, Eq.\re{Gamma-basis}, $\gamma^{(0)}_{2m+1}
= \gamma^{(0)}_{2m}=\frac12 \Gamma_{2m}^{(1)}$, $\gamma^{(1),-}_{2m-1} = -
\gamma^{(1),-}_{2m} = \Gamma_{2(m-1)}^{(2)}$ and $\gamma^{(1),+}_{2m+1} =
\gamma^{(1),+}_{2m} = \frac12\Gamma_{2(m-1)}^{(3)}$. The
$\gamma_k^{\scriptscriptstyle (0)}-$ term in \re{example} is in agreement with
findings of~\cite{Alday07}.

{\it 3.~Quantization conditions:} In our approach, the coefficients $c_p^\pm$ are
determined from the behavior of $\gamma_{2m}$ and $\gamma_{2m-1}$,
Eq.~\re{gamma=Gamma}, at large $m$. To this end, we introduce the functions
$z_\pm(x)\equiv \gamma_{2m-1}\pm \gamma_{2m}$ and examine their asymptotic
behavior in the double-scaling limit
\be\label{scaling}
m,\, g\to\infty\,,\qquad x={(m-\ft14)^2}/{g}=\mbox{fixed}\,.
\ee
Employing the approach of  \cite{Benna06} and going through numerical analysis of
$z_\pm(x)$, we found that in the limit \re{scaling} the solutions to \re{BES}
have the following remarkable scaling behavior
\ba \label{z-functions}
 z_+(x) = \frac{(gx)^{-1/4}}{g\sqrt{\pi}} \left[z_+^{(0)}(x) +
\frac{z_+^{(1)}(x)}{gx} + O(1/g^2) \right], &&
\\ \nonumber
 z_-(x) = \frac{(gx)^{-3/4}}{4g\sqrt{\pi}}  \left[z_-^{(0)}(x) +
\frac{z_-^{(1)}(x)}{gx} + O(1/g^2) \right], &&
%\\[-2mm]
\ea
where the functions $z_\pm^{(r)}(x)$ (with $r\ge 0$) do not depend on $g$ and
have faster-than-power decrease at large~$x$. For $x\to 0$, small$-x$ expansion
of $z_\pm^{(r)}(x)$ runs in {\it integer positive} powers of $x$ only.  Indeed,
matching \re{example} into \re{z-functions} we find $z_+^{(0)}(x) = 1 + c_+x +
O(x^2)$ and $z_-^{(0)}(x)=1+ (8c_--3c_+)x + O(x^2)$. For $x\to\infty$, asymptotic
behavior of $z_\pm^{(r)}(x)$ is controlled by the coefficients $c_p^\pm$. The
quantization conditions for $c_p^\pm$ follow from the requirement $\int_0^\infty
dx\,x^pz_+^{\scriptscriptstyle (r)}(x)={\rm finite}$ for any given $p\,,r\ge 0$.

Let us start with the leading term $z_\pm^{(0)}(x)$ in the expansion
\re{z-functions}. From \re{gamma=Gamma} and \re{Gamma-exact}, we evaluate
$z_\pm(x) =\gamma_{2m-1}\pm \gamma_{2m}$ in the scaling limit \re{scaling} and
find that the sums in \re{gamma=Gamma} receive dominant contribution from large
$j$. This allows us to substitute the $K_{m,j}^n-$kernel in \re{gamma=Gamma} by
its leading asymptotic behavior \re{K-asym} and evaluate sum over large $j$ in
\re{gamma=Gamma} by integration $\sum_j \mapsto \int dj$, leading to
\be
z_+(x)= -\frac{g^{-5/4}}{2\sqrt{\pi}}\sum_{p\ge 0} c_{p}^+ \Gamma(p-\ft14)
\sum_{n\ge 1} \frac{a_{n}\, x^{n+p-\ft54}}{\Gamma(n+p-\ft14)}+\ldots
\ee
where ellipses denote terms suppressed by powers of $1/g$. Then, taking the
Laplace transform w.r.t.\ $x$ we obtain
\ba\label{z+-Laplace}
&& \int_0^\infty dx \e^{-x/s} z_+(x) = -\frac{(gs)^{-1/4}}{2g\sqrt{\pi}}
\\ \nonumber
&&\qquad \times \bigg({\sum_{p\ge 0} s^{ p} c_p^+  \Gamma(p-\ft14)}\bigg)
\bigg({\sum_{n\ge 1} {a_{n} s^{n}}}\bigg)+\ldots
\ea
The sum over $n$ can be evaluated with a help of \re{a} as
\be
\sum_{n\ge 1} {a_{n} s^{n}}= -\lr{1- {\sec s}+\tan s} =-
\frac{\sqrt{2}\sin(\frac{s}2 )}{\sin(\frac{\pi}4+\frac{s}2)}\,.
\ee
As a function of $s$, it contains an infinite number of both poles and zeros on
the real $s-$axis. Requiring that the integrals $z_p\equiv \int_0^\infty dx \,
x^p z_+(x)$ should be finite for $p\ge 0$, we find that, firstly, $\int_0^\infty
dx \e^{-x/s} z_+(x)$ is an analytical function of $s$ for $\Re s > 0$ and,
secondly, it scales at large $s$ as $z_0 - z_1 /s + O(1/s^2)$. To satisfy these
conditions in the r.h.s.\ of  \re{z+-Laplace}, it proves sufficient to take
\be\label{quant1}
\sum_{p\ge 0} s^{p}\, c_p^+ \Gamma(p-\ft14) =  \xi_+ \frac{\Gamma \left( 1-
{\frac {s}{2\pi}} \right)}{\Gamma  \left( \ft34- {\frac {s}{2\pi}} \right) } +
O(1/g),
\ee
with $c_0^+=-\ft12$ and $\xi_+$ the normalization factor. Putting $s=0$ in both
sides of \re{quant1}, we get $\xi_+ = 2[\Gamma(\ft34)]^2$. Calculating the
Laplace transform $\int_0^\infty dx \e^{-x/s} z_-(x)$ in the similar manner and
imposing the same %analyticity
conditions as for $z_+(x)$ we obtain the second quantization condition
\ba\label{quant2}
 \sum_{p\ge 0} s^{p}\, \left[c_p^-\Gamma(p-\ft34) +
2c_p^+\lr{p-\ft14}{\Gamma(p+\ft14)}\right] &&
\\ \nonumber
 = {\xi_-}\frac{\Gamma \left( 1- {\frac {s}{2\pi}} \right)}{\Gamma \left( \ft14-
{\frac {s}{2\pi}} \right) }+ O(1/g), &&
\ea
with $c_0^-=0$ and $c_0^+=-\ft12$. In comparison with \re{z+-Laplace}, the
Laplace transform of $z_-(x)$ contains the factor $\sum_{n=1}^\infty {a_{n}
(-s)^{n}} = {\sqrt{2}\sin(\frac{s}2 )}/{\sin(\frac{3\pi}4+\frac{s}2)}$ that leads
to \re{quant2}. As before, putting $s=0$ in both sides of \re{quant2} we fix the
normalization factor $\xi_-=\ft14[\Gamma(\ft14)]^2$. Then, expanding both sides
of the quantization conditions \re{quant1} and \re{quant2} around $s=0$ and
matching the coefficients in front of powers of $s$, we determine the
coefficients $c_p^\pm$ (with $p\ge 1$) to the leading order in $1/g$. In this
way,
\be
c_1^+ = - {\frac {3\ln 2 }{\pi }}+\frac12+ O(1/g)\,,\quad c_1^- = {\frac {3\ln 2
}{4\pi }}-\frac14+ O(1/g).
\ee
Substituting these relations into \re{cusp-c}, we obtain %the relation for
$\Gamma_{\rm cusp}(g)$ which coincides with the string theory prediction
\re{str}.

To calculate subleading strong coupling corrections to $\Gamma_{\rm cusp}(g)$, or
equivalently to determine the coefficients $c_p^\pm$, we expand further the
Laplace transforms $\int_0^\infty dx \e^{-x/s} z_\pm(x)$ in powers of $1/g$ and
require each term of the expansion to verify the same analyticity conditions as
the leading term. This can be done systematically by applying the Euler-Maclaurin
formula to the sums over $j$ in the r.h.s.\ of \re{gamma=Gamma}. In this manner,
we obtain the following all-order quantization conditions
\ba\label{quant3}
&& \hspace*{-9mm}\sum_{p\ge 0} s^{p}\left[ c_p^+  Q^+_{p}\lr{\frac1{gs}} +
\frac1{gs} c_p^-  Q^-_{p-\frac12}\lr{\frac1{gs}}\right]
\\\nonumber
&& \hspace*{17mm} = \frac{\Gamma \left( 1- {\frac {s}{2\pi}} \right)}{\Gamma
\left( \ft34- {\frac {s}{2\pi}} \right) }\sum_{k\ge 0} (gs)^{-k} \xi^+_k(1/g),
\\\nonumber
&& \hspace*{-9mm}\sum_{p\ge 0} s^{p}\, \left[c_p^-
Q^+_{p-\frac12}\lr{\frac1{gs}}+ c_p^+  Q^-_{p}\lr{\frac1{gs}}\right]
\\\nonumber
&& \hspace*{17mm} = \frac{\Gamma \left( 1- {\frac {s}{2\pi}} \right)}{\Gamma
\left( \ft14- {\frac {s}{2\pi}} \right) } \sum_{k\ge 0} (gs)^{-k} \xi^-_k(1/g),
\ea
where $\xi^\pm_k(1/g) = \sum_{r \ge 0} \xi^\pm_{k,r} g^{-r}$ and the
($g-$independent) functions $Q^\pm_{p}(x)$ are of the form
\be
Q^+_{p}(x) = \sum_{k,\,l\ge 0} x^{k+l} Q_{k,p}^{2l}\,,\quad Q^-_{p}(x) =
\sum_{k,\,l\ge 0} x^{k+l} Q_{k,p}^{2l+1}.
\ee
Explicit expressions for the coefficients $Q_{k,p}^l$ follow univocally from the
Euler-Maclaurin summation formula and they are too lengthy to present them here.
For $g\to \infty$, the relations \re{quant3} coincide with \re{quant1} and
\re{quant2} for $\xi_0^\pm(0)=\xi_\pm$, $Q^+_{p}(0) = \Gamma(p-\ft14)$ and
$Q^-_{p}(0) =2\lr{p-\ft14} {\Gamma(p+\ft14)}$. Expanding both sides of
\re{quant3} in powers of $1/g$ and $s$ and matching the expansion coefficients,
we can determine the functions $\xi^\pm_k(g)$ and $c_p^\pm(g)$ to arbitrary order
in $1/g$. Substitution of the resulting expression for $c_p^\pm(g)$ into
\re{cusp-c} yields the strong coupling expansion of the cusp anomalous dimension.

{\it 4.~Strong coupling expansion:} Solving the quantization conditions
\re{quant3}, we calculated $\Gamma_{\rm cusp}(g)$ to order $O(1/g^{40})$. The
first few terms of the expansion
%of $\Gamma_{\rm cusp}\left(g+c_1\right)$
are
\ba\nonumber
&&{}\hspace*{-4mm} \Gamma_{\rm cusp}\left(g+c_1\right) = 2g\bigg[1- c_2 g^{-2}-
c_3 g^{-3}-\lr{ c_4+2\,c_2^{2}} g^{-4}
\\[1.2mm] \nonumber
&&{}\hspace*{-4mm}  - \lr{c_5 +23\,c_2c_3} g^{-5}- \lr{c_{{6}}+{\ft {166}{7}}
\,c_{{2}}c_{{4}} +54\,c_3^{2}+ 25\,c_2^{3}} {g}^{-6}
\\[2mm] \nonumber
&&{}\hspace*{-4mm}  - \lr{ c_{{7}}+{\ft {1721}{29}}\,c_{{2}}c_{ {5}}+{\ft
{1431}{7}}\,c_{{3}}c_{{4}}+457\, c_{{2}}^{2}c_ {{3}}} {g}^{-7}
\\[2.3mm] \nonumber
&&{}\hspace*{-4mm}   - \left(c_{{8}}+{\ft {6352}{107}}\,c_{{2}}c_{{6}}+{ \ft
{12606}{29}}\,c_{{3}}c_{{5}}+{\ft {7916}{49}} \, c_{{4}}^{2}+{\ft {6864}{7}}\,
c_{{2}}^{2}c_{ {4}} \right.
\\\label{cusp}
&&{}\hspace*{14mm} \left. +4563\,c_{{2}}c_{{3}}^{2} +374\, c_{{2}}^{4}\right)
{g}^{-8} +O \left( {g}^{-9} \right) \bigg],
\ea
where the expansion coefficients are given by
\begin{align*}
&  c_{{1}}=\frac{3\ln 2}{4\pi}, & & c_{{2}}=\frac{1}{16\pi^2}\textrm{K}, & &
c_{{3}}= \frac{{27}}{2^{11}\pi^3}\zeta(3),&
\\
& c_{{4}}=\frac{{21}}{2^{10}\pi^4}\beta(4), & &
c_{{5}}=\frac{{43065}}{2^{21}\pi^5}\zeta(5) ,& & c_{{6}}=
\frac{{1605}}{2^{15}\pi^6}\beta(6), &
\end{align*}\\[-10.46mm]
%\\
%&  c_{{7}}=  \frac{{101303055}}{2^{30}\pi^7}\zeta(7), && c_{{8}}=
%\frac{{1317645}}{2^{22}\pi^8}\beta(8), && \ \ \ldots &
%\end{align*}
\be\label{cc}
c_{{7}}=  \frac{{101303055}}{2^{30}\pi^7}\zeta(7), \qquad c_{{8}}=
\frac{{1317645}}{2^{22}\pi^8}\beta(8), %\ \ldots
\ee
with $\zeta(x)$ %=\sum_{n\ge 1}{n^{-x}}$
the Riemann zeta function, $\beta(x)=\sum_{n\ge 0}{(-1)^n}{ (2n+1)^{-x}}$ the
Dirichlet beta function and $\textrm{K}=\beta(2)$ the Catalan's constant. We
verified that the coefficients \re{cc} are in an excellent agreement with the
numerical values obtained within the approach of \cite{Benna06}.

The following concluding remarks are in order.

The reason why in \re{cusp} we expanded $\Gamma_{\rm cusp}\left(g+c_1\right)$
instead of $\Gamma_{\rm cusp}\left(g\right)$ is that the $c_1-$dependent terms
inside $\Gamma_{\rm cusp}(g)$ can be resummed to all orders in $1/g$ by simply
replacing $g \to g + c_1$.
%According to \re{cusp}, the $c_1-$dependent terms inside $\Gamma_{\rm cusp}(g)$
%can be resummed to all orders in $1/g$ by simply replacing $g \to g + c_1$.
This suggests to use $g'\equiv g-c_1$ as a new parameter of the strong coupling
expansion.

A distinguished feature of the series \re{cusp} is that the coefficients in front
of $1/g^n$ are given by a linear combination of the product of $\zeta(2p+1)$ and
$\beta(2r)$ such that sum of their arguments equals $n$. Let us compare this with
the weak coupling expansion of $\Gamma_{\rm cusp}(g)$. The latter runs in even
powers of $g$ and the expansion coefficients only involve products of
$\zeta-$functions of both even and odd arguments such that the sum of their
arguments equals the order in $g$~\cite{BES06,KL07}.

We found that, up to order $O(1/g^{40})$, all expansion coefficients of
$\Gamma_{\rm cusp}(g)$ except the first one are negative. In addition, at large
orders in $1/g$, they grow factorially and the asymptotic expansion is not Borel
summable
\be
\Gamma_{\rm cusp}(g)\! \sim \! -g \sum_k \frac{\Gamma(k-\ft12)}{(2\pi g)^k} = g
\int_0^\infty \frac{du\, u^{-1/2}\e^{-u}}{u-2\pi g},
\ee
with the Stieltjes integral having a pole at $u=2\pi g$. This indicates that
$\Gamma_{\rm cusp}(g)$ receives nonperturbative correction $\sim g^{1/2}
\e^{-2\pi g}$ proportional to the residue at the pole.

Our prediction for the cusp anomalous dimension \re{cusp} relies on the strong
coupling expansion of the solution to the BES equation \re{BES}. Eventual
verification of \re{cusp} remains a challenge for the string theory. We would
like to mention that our result for $c_2={\rm K}/(4\pi)^2$ is in a structural
agreement with the (revised) two-loop superstring result of \cite{RTT07} and in
precise agreement with a new superstring computation (R.~Roiban  and
A.A.~Tseytlin, arXiv:0709.0681 [hep-th]).

We are most grateful to A.~Gorsky for collaboration at an early stage of the
project. We would like to thank A.~Belitsky, J.~Maldacena, A.~Manashov,
A.~Tseytlin and A.~Turbiner for interesting discussions and correspondence. This
research has been supported in part by the French Agence Nationale de la
Recherche, contract ANR-06-BLAN-0142 (B.B., G.K.), by the Foundation for Polish
Science and by the MEiN grant P03B-024-27 (J.K.).

{\it Note added:} After this paper was submitted to Arxiv, we learned
(J.~Maldacena, private communication) that our result for nonperturbative
corrections to $\Gamma_{\rm cusp}(g)$ is in a perfect agreement with the findings
of Ref.~\cite{AM07}. As was shown in \cite{AM07}, $\Gamma_{\rm cusp}(g)$, has the
interpretation of an energy density of a certain flux configuration and, as such,
it receives correction proportional to $m^2$ with $m \sim g^{1/4} \e^{-\pi g}$
being the mass gap in the $O(6)$ sigma model.

\end{document}